\documentclass[11pt]{article}

\usepackage[affil-it]{authblk}
\usepackage{algorithm}
\usepackage{algpseudocode}
\usepackage{subcaption}
\usepackage{graphicx}
\usepackage{amssymb,amsmath}
\usepackage{mathtools}
\usepackage{accents}
\usepackage{bbm}
\usepackage{hyperref}
\usepackage[all]{hypcap} 
\hypersetup{
  colorlinks   = true,
  urlcolor     = blue,
  linkcolor    = red,
  citecolor   = blue
}
\usepackage{natbib}

\newcommand{\condindep}{\perp\!\!\!\perp}

\begin{document}
\title{Bayesian test of significance for conditional independence: The multinomial model}
\author{Pablo M. Andrade\thanks{Email: \texttt{pablo.andrade@usp.br}},
		~Julio M. Stern \thanks{Email: \texttt{jstern@ime.usp.br}}
		~and Carlos Alberto de Bragan\c{c}a Pereira \thanks{Email: \texttt{cpereira@ime.usp.br}}}
\affil{Instituto de Matem\'atica e Estat\'{\i}stica, \\
	   Universidade de S\~ao Paulo (IME-USP)\\
	   Rua do Mat\~ao, 1010, Cidade Universit\'aria,\\
	   S\~ao Paulo, SP/Brasil, CEP: 05508-090}
\maketitle

\begin{abstract}
Conditional independence tests (CI tests) have received special attention lately in Machine Learning and Computational Intelligence related literature as an important indicator of the relationship among the variables used by their models. In the field of Probabilistic Graphical Models (PGM)--which includes Bayesian Networks (BN) models--CI tests are especially important for the task of learning the PGM structure from data. In this paper, we propose the Full Bayesian Significance Test (FBST) for tests of conditional independence for discrete datasets. FBST is a powerful Bayesian test for precise hypothesis, as an alternative to frequentist's significance tests (characterized by the calculation of the \emph{p-value}).
\end{abstract}

\section{Introduction}
\label{sec:intro}
\citet{RefBarlowPereira90} discuss a graphical approach to conditional independence. A probabilistic influence diagram is a directed acyclic graph (DAG) that helps to model statistical problems. The graph is composed of a set of nodes or vertices, representing the variables, and a set of arcs joining the nodes, representing the dependence relationships shared by these variables.

The construction of the model helps to understand the problem and gives a good representation of interdependence of the variables involved in the problem. The joint probability of these variable can be written as a product of conditional distributions, based on the relationships of independence and conditional independence among the variables involved in the problem.

Sometimes the interdependence of the variables is not known, and in this case, the model structure is required to be learnt from data. Algorithms such as the \emph{IC-Algorithm (Inferred Causation)} described in \citet{RefPearlVerma95} are designed to uncover these structures from data. This algorithm uses a series of CI tests to remove and direct the arcs connecting the variables in the model, returning a DAG that minimally (with the minimum number of parameters, without loss of information) represents the variables in the problem.

The problem of learning DAG structures from data motivates the proposal of new powerful statistical tests for the hypothesis of conditional independence, since the accuracy of structures learnt are directly affected by errors committed by these tests. Recently proposed structure learning algorithms \citep[see][]{RefChengJ97,RefTsamardinos06,RefYehezkelLerner09} indicate as main source of errors the results of CI tests.

In this paper, we propose the Full Bayesian Significance Test (FBST) for tests of conditional independence for discrete datasets. FBST is a powerful Bayesian test for precise hypothesis, and can be used to learn DAG structures from data, as an alternative to CI test currently used, such as \emph{Pearson's} $\chi^2$ \emph{test}.

This paper is organized as follows. In Section~\ref{sec:fbst} we review the Full Bayesian Significance Test (FBST). In Section~\ref{sec:fbstcomposition}, we review the FBST for composite hypothesis. Section~\ref{sec:CItestContTable} shows an example of test of conditional independence used to learn a simple model with 3 variables.

\section{The Full Bayesian Significance Test}
\label{sec:fbst}
The Full Bayesian Significance Test (FBST) is presented by \citet{RefPereiraStern99} as a coherent Bayesian significance test for sharp hypothesis. In the FBST, the evidence for a precise hypothesis is computed.

This evidence is given by the complement of the probability of a credible set--called the \emph{tangent} set--which is a subset of the parameter space, where the posterior density of each of its elements is greater than the maximum of the posterior density over the Null hypothesis. A more formal definition is given below.

Consider a model in a statistical space described by the triple $\left(\Xi, \Delta, \Theta \right)$, where $\Xi$ is the sample space; $\Delta$, the family of measurable subsets of $\Xi$; and $\Theta$ the parameter space: $\Theta$ is a subset of $\Re^n$.

Define a subset of the parameter space $T_{\varphi}$ (\emph{tangent} set), where the posterior density (denoted by $f_x$) of each element of this set is greater than $\varphi$.
\begin{align*}
T_{\varphi} = \left\{ \theta \in \Theta | f_x(\theta)>\varphi \right\}
\end{align*}
The credibility of $T_{\varphi}$ is given by its posterior probability:
\begin{align*}
\kappa = \int_{T_{\varphi}}f_x(\theta)d \theta = \int_{\Theta}f_x(\theta)\mathbbm{1}_{T_{\varphi}}\left(\theta \right) d \theta
\end{align*}
, where $\mathbbm{1}_{T_{\varphi}}\left ( \theta \right)$ is the indicator function:
\[
\mathbbm{1}_{T_{\varphi}}\left(\theta\right) = \begin{cases} 
        1 & \text{if } \theta \in T_{\varphi}\\
        0 & \text{otherwise}
		 \end{cases}
\]
Defining the maximum of the posterior density over the Null hypothesis as $f^*_x$, with  maximum point at $\theta_0^*$:
\begin{align*}
\theta_0^* \in \underset{\theta\in\Theta_0}{\operatorname{argmax}}f_x\left(\theta\right),\text{ and } f_x^*=f_x(\theta^*)
\end{align*}
, and defining $T^*=T_{f_x^*}$ the tangent set to the Null hypothesis $H_0$. The credibility of $T^*$ is $\kappa^*$

The measure of evidence of the Null hypothesis (called \emph{e-value}), which is the complement of the probability of the set $T^*$, is defined as:
\begin{align*}
Ev(H_0)=1-\kappa^*=1-\int_{\Theta}f_x(\theta)\mathbbm{1}_{T^*}\left(\theta \right) d \theta
\end{align*}
If the probability of the set $T^*$ is large, the null set is in a region of low probability and the evidence is against the Null hypothesis $H_0$. But, if the probability of $T^*$ is small, then the null set is in a region of high probability, and the evidence supports the Null hypothesis.
\subsection{FBST: Example of Tangent set}
\label{sec:fbstexampleTset}
Figure~\ref{fig:tangetSet} shows the tangent set for a Null hypothesis $H_0: \mu=1$, for the posterior distribution $f_x$ given bellow, where $\mu$ is the mean of a normal distribution and $\tau$, the \textit{precision} (the inverse of the variance $\tau=\frac{1}{\sigma^2}$): 
\begin{align*}
f_x(\mu,\tau) \propto \tau^{1.5}e^{-\tau (\mu)^2-1.5\tau}
\end{align*}
\section{FBST: Compositionality}
\label{sec:fbstcomposition}
The relationship between the credibility of a complex hypothesis $H$, and its elementary constituent, $H_j$, $j=1,\ldots,k$, under the Full Bayesian Significance Test (FBST), is analysed in \citet{RefBorgesStern07}.

For a given set of \emph{independent} parameters $\left( \theta_1,\ldots,\theta_k \right) \in \left(\Theta_1\times\ldots\times\Theta_k \right)$, a complex hypothesis $H$, such as:
\begin{figure}[H]
	\begin{minipage}[b]{0.49\linewidth}
	\centering
		\includegraphics[scale=0.40]{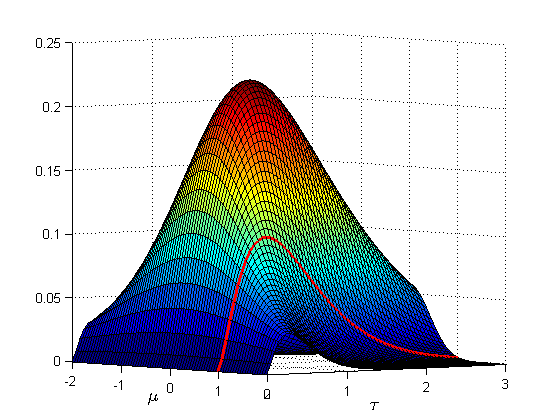}
		\subcaption{Posterior $f_x$. Red line: $\mu=1.0$.\label{fig:tangetSet_a}}
	\end{minipage}
	\begin{minipage}[b]{0.49\linewidth}
	\centering
		\includegraphics[scale=0.40]{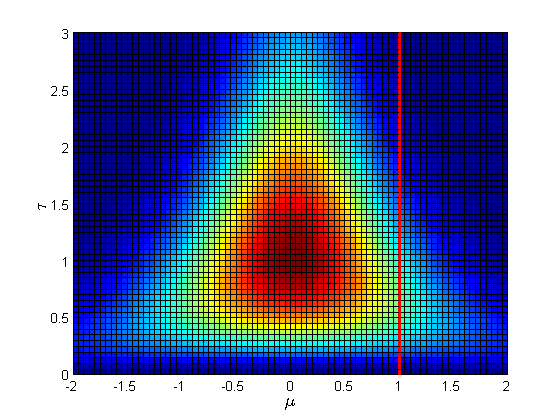}
		\subcaption{Posterior $f_x$. Red line: $\mu=1.0$.\label{fig:tangetSet_b}}
	\end{minipage}
	\begin{minipage}[b]{1.0\linewidth}
	\centering
		\includegraphics[scale=0.50]{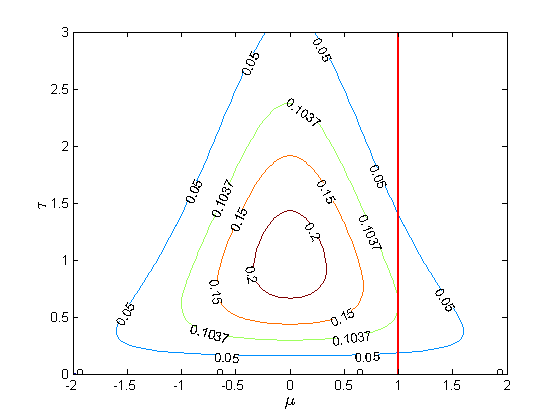}
		\subcaption{Contours of $f_x$. Red line: $\mu=1.0$.\label{fig:tangetSet_c}}
	\end{minipage}
	\caption{Example of tangent set for a Null hypothesis $H_0: \mu=1.0$. In (a) and (b) the posterior
	distribution $f_x$ is shown, with the red line representing the points in the Null hypothesis
	($\mu=1$). In (c) the contours of $f_x$ show that the points of maximum density in the Null
	hypothesis $\theta^*_0$ have density 0.1037 ($f^*=f\left(\theta_0^*\right)=0.1037$). The tangent set
	$T^*$ of the Null hypothesis $H_0$ is the set of points inside the green contour line (points with
	density greater than $f^*$), and the e-value of $H_0$ is the complement of the integral of $f_x$
	bounded by the green contour line.\label{fig:tangetSet}}
\end{figure}
\begin{align*}
H: \theta_1 \in \Theta_1^{H} \wedge \theta_2 \in \Theta_2^{H} \wedge \ldots \wedge \theta_k^{H} \in \Theta_k^{H}
\end{align*}
, where $\Theta_j^H$ is a subset of the parameter space $\Theta_j$ for $j=1,\ldots,k$, constrained to the hypothesis $H$, can be decomposed in its elementary components (hypotheses):
\begin{gather*}
H_1: \theta_1 \in \Theta^H_1 \\
H_2: \theta_2 \in \Theta^H_2 \\
\cdots \\
H_k: \theta_k \in \Theta^H_k
\end{gather*}
, and the credibility of $H$ can be evaluated based on the credibility of these components. The evidence in favour of the complex hypothesis $H$ (measured by its \emph{e-value}) can not be obtained directly from the evidence in favour of the elementary components, but based on their \emph{Truth Function} $W^j$ (or cumulative surprise distribution) defined below.

For a given elementary component ($H_j$) of the complex hypothesis $H$, $\theta_{j}^*$ is the point of maximum density of the posterior distribution ($f_x$) constrained to the subset of the parameter space defined by hypothesis $H_j$:
\begin{align*}
\theta_{j}^* \in \underset{\theta_{j}\in \Theta^H_{j}}{\text{argmax}}f_x\left(\theta_{j}\right) \text{ and } f_{j}^{*}=f_x\left(\theta_j^{*}\right)
\end{align*}
The \emph{truth function} $W_j$ is the probability of the region of the parameter space, where the posterior density is lower or equal than a value $f$:
\begin{align*}
R_j(f) = \{\theta_j \in \Theta_j|f_x\left(\theta_j\right)\leq f\} \\
W_j(f) = \int_{R_j(f)} f_x\left(\theta_j\right)d\theta_j
\end{align*}
And the evidence supporting the hypothesis $H_j$ is:
\begin{align*}
Ev(H_j)=W_j(f_j^{*})
\end{align*}
The evidence supporting the complex hypothesis can be then described in terms of the \emph{truth function} of its components, as the Mellin convolution of these functions:
\begin{align*}
Ev(H)=W_1\otimes W_2\otimes W_3\otimes \ldots \otimes W_k\left(f_1^{*}\cdot f_2^{*} \cdot f_3^{*} \cdot \ldots \cdot f_k^{*}\right)
\end{align*}
Where the Mellin Convolution of two \emph{truth functions}, $W_1\otimes W_2$, is the distribution function:
\begin{align*}
W_1 \otimes W_2(x) = \int_0^{x} W_1\left(\frac{x}{y}\right)W_2(y)dy
\end{align*}
\subsection{Numerical Method for Convolution and Condensation}
\label{sec:convcondens}
\citet{RefWilliamsonDowns90} investigate numerical procedures to handle arithmetic operations for random variables. Replacing basic operations of arithmetic, used for fixed numbers, by convolutions, they show how to calculate the joint distribution for a set of random variables and their respective upper and lower bounds.

The convolution for the multiplication of two random variables $X_1$ and $X_2$ ($Z=X_1 \cdot X_2$) can be written using their respective cumulative distribution functions $F_{X_1}$ and $F_{Y_2}$:
\begin{align*}
F_Z(z)=\int_{0}^z F_{X_1}\left(\frac{z}{t}\right)dF_{X_2}(t)
\end{align*}
The algorithm for the numerical calculation of the distribution of the product of two independent random variables ($Y_1$ and $Y_2$), using their \emph{discretized} marginal probability distributions ($f_{Y_1}$ and $f_{Y_2}$) is shown in Algorithm~\ref{alg:convolution} (an algorithm for a discretization procedure is given in \citealt*[page 188]{RefWilliamsonDowns90}).

The numerical convolution of two distributions with $N$ bins returns a distribution with $N^2$ bins. For a sequence of operations, this would be a problem, since the result of each operation would be larger than the input for the operations. The authors, hence, propose a simple method to reduce the size of the output to $N$ bins, without introducing error to the result. This operation is called \emph{condensation}, and it returns the upper and lower bounds of each of the $N$ bins for the distribution resulting from the convolution. The algorithm for the condensation process is shown in Algorithm~\ref{alg:condensation}.

\begin{algorithm}[H]
	\caption{Find distribution of the product of two random variables.\label{alg:convolution}}
	\begin{algorithmic}[1]
		\Procedure{Convolution}{$f_{Y_1},f_{Y_2}$}\Comment{Discrete p.d.f. of $Y_1$ and $Y_2$}
			\State $f \gets array(0,size \gets n^2)$ 		\Comment{$f$ and $W$ has $n^2$ bins}
   			\State $W \gets array(0,size \gets n^2)$ 		
   			\For{$i \gets 1,n$} \Comment{$f_1$ and $f_2$ have $n$ bins}
	   			\For{$j \gets 1,n$} 
  	    			\State $f[(i-1)\cdot n+j] \gets f_{Y_1}[i]\cdot f_{Y_2}[j]$
	   			\EndFor
   			\EndFor
   			\State $W[1] \gets f[1]$
   			\For{$i \gets k,n^2$} \Comment{find c.d.f. of $Y_1\cdot Y_2$}
   				\State $W[k] \gets f[k]$
   				\State $W[k] \gets W[k]+ W[k-1]$
	   		\EndFor
   			\State \textbf{return} $W$ \Comment{Discrete c.d.f. of $Y_1\cdot Y_2$}
		\EndProcedure
	\end{algorithmic}
\end{algorithm}

\begin{algorithm}[H]
	\caption{Find upper lower bound for a c.d.f. for condensation.\label{alg:condensation}}
	\begin{algorithmic}[1]
		\Procedure{HorizontalCondensation}{$W$} \Comment{Histogram of a c.d.f. with $n^2$ bins}
   			\State $W^l \gets array(0,size \gets n)$ 
   			\State $W^u \gets array(0,size \gets n)$ 		
   			\For{$i\gets 1,n$}
  	    		\State $W^l[i] \gets W[(i-1)\cdot n+1]$ \Comment{lower bound after condensation}
  	    		\State $W^u[i] \gets W[i \cdot n]$ \Comment{upper bound after condensation}
   			\EndFor
   			\State \textbf{return} $\left[W^l,W^u\right]$\Comment{Histograms with upper/lower bounds}
		\EndProcedure
	\end{algorithmic}
\end{algorithm}
\subsubsection{Vertical Condensation}
\citet{RefKaplanLin87} propose a \emph{vertical} condensation procedure for discrete probability calculations, where the condensation is done using the vertical axis, instead of the horizontal axis, as in \citet{RefWilliamsonDowns90}.

The advantage of this approach is that it provides more control over the representation of the distribution, since, instead of selecting an interval of the domain of the cumulative distribution function (values assumed by the random variable) as a bin, we select the interval of the range of the cumulative distribution in $[0,1]$ that should be represented by each bin. 

In this case, it is also possible to concentrate the attention in a specific region of the distribution. For example, if there is a greater interest in the behaviour of the tail of the distribution, the size of the bins can be reduced in this region, consequently, increasing the number of bins necessary to represent the tail of the distribution.

An example of convolution followed by condensation procedure, using both approaches is given in Section~\ref{sec:mellinExample}. We used, for this example, discretization and condensation procedures with bins \emph{uniformly} distributed over both axes. At the end of the condensation procedure, using the first approach, the bins are uniformly distributed \emph{horizontally} (over the sample space of the variable). For the second approach, the bins of the cumulative probability distribution are uniformly distributed over the vertical axis in the interval $[0,1]$. Algorithm~\ref{alg:verticalcondensation} shows the condensation with bins uniformly distributed over the vertical axis.
\begin{algorithm}[H]
	\caption{Condensation with bins vertically uniformly distributed.\label{alg:verticalcondensation}}
	\begin{algorithmic}[1]
		\Procedure{VerticalCondensation}{$W$,$f$,$x$} 
		\Comment{Histograms of a c.d.f. and p.d.f., and breaks in the x axis.}
			\State $breaks \gets \left[1/n,2/n,...,1\right]$ \Comment{uniform breaks in $y$ axis}
			\State $W_n \gets array\left(0,size \gets n\right]$
			\State $x_n \gets array\left(0,size \gets n\right]$
			\State $lastbreak \gets 1$			
			\State $i \gets 1$
			\ForAll{$b \in breaks$}
				\State $w \gets first(W \geq b)$ \Comment{find break to create current bin}
				\If{$W[w] \neq b$} \Comment{if the break is within a current bin}
					\State $ratio \gets (b-W[w-1])/(W[w]-W[w-1])$
					\State $x_n[i] \gets \frac{1}{1/n} \left(sum \left(f[w-1]\cdot x[w-1]\right)+ratio \cdot f[w] \cdot x[w]\right)$
					\State $W[i-1] \gets b$
					\State $W_n[i] \gets b$
					\State $f[i-1] \gets f[w-1]+ratio \cdot f[w]$
					\State $f[i] \gets (1-ratio)\cdot f[w]$
				\Else
					\State $x_n[i] \gets x[w]$
					\State $W_n[i] \gets W[w]$
				\EndIf
				\State $lastbreak \gets b$
				\State $i \gets i+1$
			\EndFor
   			\State \textbf{return} $\left[W_n,x_n\right]$\Comment{Histograms with upper/lower bounds}
		\EndProcedure
	\end{algorithmic}
\end{algorithm}

\subsection{Mellin Convolution: Example}
\label{sec:mellinExample}
An example of Mellin convolution to find the product of two random variable $Y_1$ and $Y_2$, both with a Log-normal distribution, is given.\\\\
Assume $Y_1$ and $Y_2$, continuous random variables, such that.
\begin{align*}
Y_1 \sim \ln \mathcal{N}\left(\mu_1,\sigma_1^2\right) \text{, and } Y_2 \sim \ln \mathcal{N}\left(\mu_2,\sigma_2^2\right)
\end{align*}
, we denote the cumulative distributions of $Y_1$ and $Y_2$, by $W_1$ and $W_2$, respectively, i.e.,
\begin{align*}
W_1(y_1) = \int_{-\infty}^{y_1} f_{Y_1}(t)d t \text{, and } W_2(y_2) = \int_{-\infty}^{y_2} f_{Y_2}(t)d t
\end{align*}
, where $f_{Y_1}$ and $f_{Y_2}$ are the density functions of $Y_1$ and $Y_2$, respectively. These distributions can be written as a function of two normally distributed random variables $X_1$ and $X_2$:
\begin{align*}
\ln(Y_1)=X_1 \sim \mathcal{N}\left(\mu_1,\sigma_1^2\right) \\
\ln(Y_2)=X_2 \sim \mathcal{N}\left(\mu_2,\sigma_2^2\right)
\end{align*}

And we can find the distribution of the product of these random variables ($Y_1 \cdot Y_2$), using simple arithmetic operations, to be also Log-normal:  
\begin{align*}
Y_1=e^{X_1} \text{ and } Y_2=e^{X_2} \\
Y_1 \cdot Y_2 =e^{X_1+X_2} \\
\ln(Y_1 \cdot Y_2)=X_1+X_2 \sim \mathcal{N}\left(\mu_1+\mu_2,\sigma_1^2+\sigma_2^2\right) \\
\therefore Y_1 \cdot Y_2 \sim \ln \mathcal{N}\left(\mu_1+\mu_2,\sigma_1^2+\sigma_2^2\right) \\
\end{align*}

The cumulative density function of $Y_1 \cdot Y_2$ ($W_{12}(y_{12})$) is defined as:
\begin{align*}
W_{12}(y_{12}) = \int_{-\infty}^{y_{12}} f_{Y_1 \cdot Y_2}(t)d t
\end{align*}
, where $f_{Y_1\cdot Y_2}$ is the density function of $Y_1\cdot Y_2$.

\begin{figure}[h]
	\begin{minipage}[b]{0.49\linewidth}
	\centering
		\includegraphics[scale=0.30]{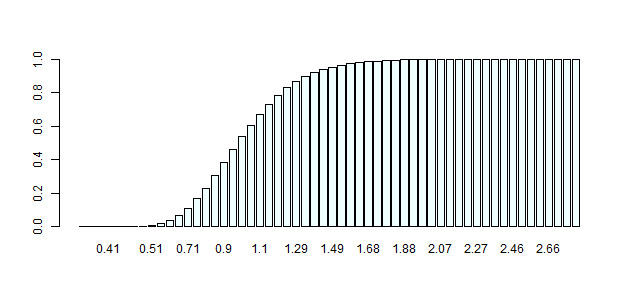}
		\subcaption{$W_1$: Horinzontal discretization\label{fig:cumhisthy1}}
	\end{minipage}
	\begin{minipage}[b]{0.49\linewidth}
	\centering
		\includegraphics[scale=0.30]{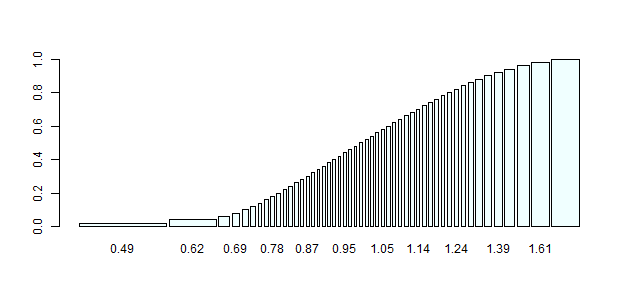}
		\subcaption{$W_1$: Vertical discretization\label{fig:cumhisthy1v}}
	\end{minipage}\\
	\begin{minipage}[b]{0.49\linewidth}
	\centering
		\includegraphics[scale=0.30]{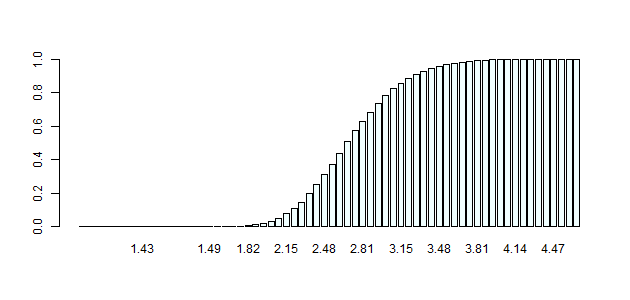}
		\subcaption{$W_2$: Horinzontal discretization\label{fig:cumhisthy2}}
	\end{minipage}
	\begin{minipage}[b]{0.49\linewidth}
	\centering
		\includegraphics[scale=0.30]{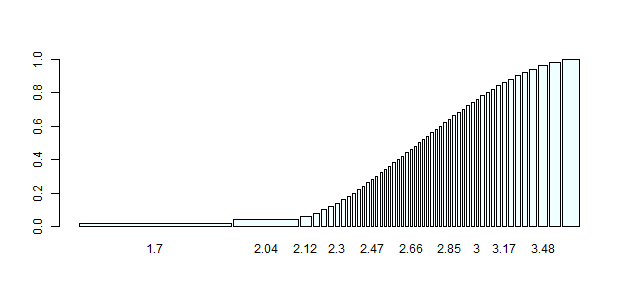}
		\subcaption{$W_2$: Vertical discretization\label{fig:cumhisthy2v}}
	\end{minipage}\\
	\caption{Example of different discretization methods for the representation of the c.d.f. of two random variables
	($Y_1$ and $Y_2$) with Log-normal distribution. In (a) and (c) the c.d.f. of $Y_1$ and $Y_2$, respectively, with
	bins uniformly distributed over the x-axis are shown, in (b) and (d) the c.d.f. of $Y_1$ and $Y_2$, respectively,
	with bins uniformly distributed over the y-axis.\label{fig:exdiscretization}}
\end{figure}

\begin{figure}[h]
	\begin{minipage}[b]{0.49\linewidth}
	\centering
		\includegraphics[scale=0.30]{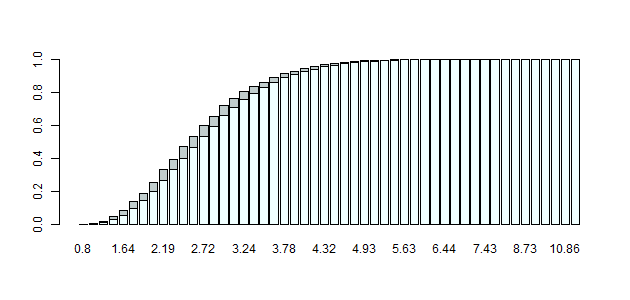}
		\subcaption{$W_1 \otimes W_2$: Horizontal condensation\label{fig:convolutionw1w2}}
	\end{minipage}
	\begin{minipage}[b]{0.49\linewidth}
	\centering
		\includegraphics[scale=0.30]{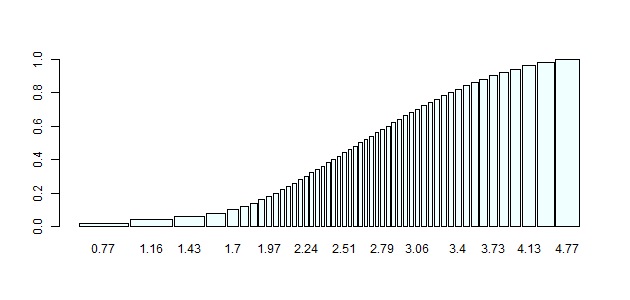}
		\subcaption{$W_1 \otimes W_2$: Vertical condensation\label{fig:convolutionw1w2V}}
	\end{minipage}\\
	\begin{minipage}[b]{0.49\linewidth}
	\centering
		\includegraphics[scale=0.30]{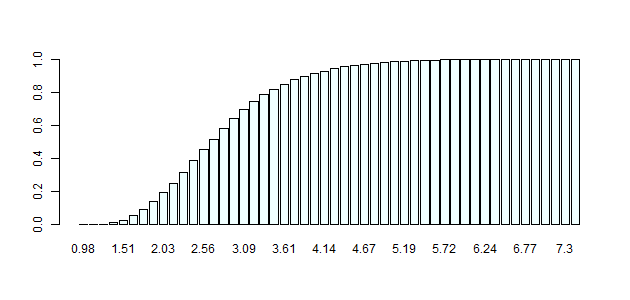}
		\subcaption{$Y_1\cdot Y_2$: Horizontal discretizarion\label{fig:cumhisthy1timesy2}}
	\end{minipage}%
	\begin{minipage}[b]{0.49\linewidth}
	\centering
		\includegraphics[scale=0.30]{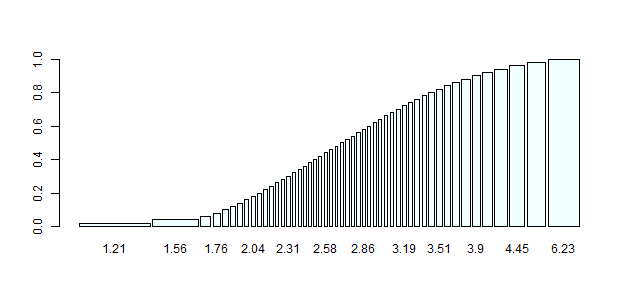}
		\subcaption{$Y_1\cdot Y_2$: Vertical discretizarion\label{fig:cumhisthy1timesy2V}}
	\end{minipage}
	\caption{Example of convolution of two random variables ($Y_1$ and $Y_2$) with Log-normal distribution.
	The result of the convolution $Y_1 \otimes Y_2$, followed by horizontal condensation (bins uniformly distributed
	over x-axis) is shown in (a), and by vertical condensation (bins uniformly distributed
	over y-axis) is shown in (b). The true distribution of the product $Y_1 \cdot Y_2$ is shown in
	(c) and (d), respectively, for horizontal and vertical discretization procedures.\label{fig:exconvolution}}
\end{figure}

Figure~\ref{fig:exdiscretization} shows the cumulative distribution functions of $Y_1$ and $Y_2$ discretized with bins uniformly distributed over both x and y axes (horizontal and vertical discretizations). Figure~\ref{fig:exconvolution} shows an example of convolution followed by condensation, using both horizontal and vertical condensation procedures, and the true distribution of the product of two variables with Log-normal distributions.

\section{Test of Conditional Independence in Contingency table using FBST}
\label{sec:CItestContTable}
We now apply the methods shown in the previous sections to find the evidence of a complex Null hypothesis of conditional independence, for discrete variables.

Given the discrete random variables $X$, $Y$ and $Z$, with $X$ taking values in $\{1,\ldots,k\}$. The test of conditional independence $Y \condindep Z|X$ can be written as the complex Null hypothesis $H$:
\begin{align*}
H: [Y \condindep Z |X=1] \wedge [Y \condindep Z |X=2] \wedge  \cdots \wedge [Y \condindep Z |X=k]
\end{align*}
The hypothesis $H$, can be decomposed in its elementary components:
\begin{gather*}
H_1: Y \condindep Z |X=1\\
H_2: Y \condindep Z |X=2\\
\cdots \\
H_k: Y \condindep Z |X=k
\end{gather*}
Notice that the hypotheses $H_1,\ldots,H_k$ are \emph{independent}: for each value $x$ taken by $X$, the values taken by variables $Y$ and $Z$ are assumed to be random observations drawn from some distribution $p(Y,Z|X=x)$. Each of the elementary components is a hypothesis of independence in a contingency table. Table~\ref{tab:conttable} shows the contingency table for $Y$ and $Z$ taking values, respectively, in $\{1,\ldots,r\}$ and $\{1,\ldots,c\}$.
\begin{table}[h]
	\centering
	\caption{Contingency table of $Y$ and $Z$ for $X=x$ (hypothesis $H_x$): $n_{yzx}$ is the count of 
	$[Y,Z]=[y,z]$, when $X=x$.\label{tab:conttable}}
	\begin{tabular}{l || c | c | c | c }
		\hline\noalign{\smallskip}
				 & $Z=1$ &  $Z=2$ & $\cdots$ & $Z=c$\\
			\noalign{\smallskip}\hline\noalign{\smallskip}
		$Y=1$	 & $n_{11x}$ & $n_{12x}$ & $\cdots$ & $n_{1cx}$\\
		$Y=2$	 & $n_{21x}$ & $n_{22x}$ & $\cdots$ & $n_{2cx}$\\
		$\cdots$ & $\cdots$ & $\cdots$ & $\cdots$ & $\cdots$\\
		$Y=r$	 & $n_{r1x}$ & $n_{r2x}$ & $\cdots$ & $n_{rcx}$\\
		\noalign{\smallskip}\hline
	\end{tabular}
\end{table}
The test of the hypothesis $H_x$ can be set-up using the multinomial distribution for the cell counts of the contingency table and its natural conjugate prior, the Dirichlet distribution for the vector of parameters $\theta_x=\left[\theta_{11x},\theta_{12x},\ldots,\theta_{rcx}\right]$.

For a given array of hyperparameters $\alpha_{x}=[\alpha_{11x},\ldots,\alpha_{rcx}]$, the Dirichlet distribution is defined as:
\begin{align}
\label{eq:priordist}
f\left(\theta_x|\alpha_x\right)=
\Gamma\left(\sum_{y,z}^{r,c}\alpha_{yzx}\right)
\prod_{y,z}^{r,c} \frac{\theta_{yzx}^{\alpha_{yzx}-1}}{\Gamma\left(\alpha_{yzx}\right)} 
\end{align}
The multinomial likelihood, for the given contingency table, assuming the array of observations $n_x=[n_{11x},\ldots,n_{rcx}]$ and the sum of the observations $n_{..x}=\sum_{y,z}^{r,c}n_{yzx}$, is:
\begin{align}
\label{eq:multinomialdist}
f\left(n_x|\theta_x\right)=
n_{..x}!\prod_{y,z}^{r,c} \frac{\theta_{yzx}^{n_{yzx}}}{n_{yzx}!}
\end{align}
The posterior distribution will be, then, a Dirichlet distribution $f_{n}(\theta_x)$:
\begin{align}
\label{eq:posteriordist}
f_n\left(\theta_x\right) \propto
\prod_{y,z}^{r,c}\theta_{yzx}^{\alpha_{yzx}+n_{yzx}-1}
\end{align}
Under the hypothesis $H_x$, we have $Y \condindep Z | X=x$. In this case, we have that the joint distribution is equal to the product of the marginals: $p\left(Y=y,Z=z|X=x\right)=p\left(Y=y|X=x\right)p\left(Z=z|X=x\right)$. We can define this condition using the array of parameters $\theta_x$, in this case, we have:
\begin{align}
\label{eq:pointmaxdensityH}
H_x: \theta_{yzx}= \theta_{.zx}\cdot \theta_{y.x}, \forall y,z
\end{align}
, where $\theta_{.zx}=\sum_{y}^{r}n_{yzx}$ and $\theta_{y.x}=\sum_{z}^{c}\theta_{yzx}$.

The point of maximum density of the posterior distribution constrained to the subset of the parameter space defined by the hypothesis $H_x$ can be estimated using the \emph{maximum a posteriori} (MAP) estimator under the hypothesis $H_x$ (the mode of the parameters $\theta_x$). The maximum density ($f_x^*$) will be the posterior density evaluated at this point:
\begin{align}
\label{eq:maxdensityH}
\theta^*_{yzx} = \frac{n^{H_x}_{yzx}+\alpha_{yzx}-1}{n^{H_x}_{..x}+\alpha_{..x}- r\cdot c} \text{ and }
f_x^*=f_n(\theta_x^*)
\end{align}
, where $\theta_x^*=\left[\theta^*_{11x},\ldots,\theta^*_{rcx}\right]$.

The evidence supporting $H_x$ can be written in terms of the \emph{truth function} $W_x$, as defined in Section~\ref{sec:fbstcomposition}:
\begin{align}
\label{eq:regiontruthfunc}
R_x(f) = \{\theta_x \in \Theta_x|f_x\left(\theta_x\right)\leq f\}\\
W_x(f) = \int_{R_x(f)}f_n\left(\theta_x\right)d\theta_x \propto \int_{R_x(f)}\prod_{y,z}^{r,c}\theta_{yzx}^{\alpha_{yzx}+n_{yzx}-1}d\theta_x
\end{align}
And the evidence supporting the hypothesis $H_x$, is:
\begin{align}
\label{eq:evalue}
Ev(H_x)=W_x(f_{x}^{*})
\end{align}
Finally the evidence supporting the hypothesis of conditional independence ($H$), will be given by the convolution of the \emph{truth functions} evaluated at the product of the points of maximum posterior density, for each component of the hypothesis $H$:
\begin{align}
\label{eq:Hevalue}
Ev(H) = W_1 \otimes W_2 \otimes \ldots \otimes W_k \left( f^*_1 \cdot f^*_2 \cdot \ldots \cdot f^*_k\right)
\end{align}
The \emph{e-value} for hypothesis $H$ can be found using modern mathematical methods of integration. An example is given in the next section, where the numerical convolution followed by the condensation procedures described in Section~\ref{sec:convcondens} are used. The application of the method of horizontal condensation results in a interval for the e-value (found using the lower and upper bounds resulting from the condensation process), and in a single value for the vertical procedure.

\subsection{Example of CI test using FBST}
In this section we describe an example of CI test using the Full Bayesian Significance Test (FBST) for conditional independence using samples from two different model. For both models, we test if the variable $Y$ is conditionally independent of $Z$ given $X$.
\begin{figure}[H]
	\begin{minipage}[b]{0.49\linewidth}
		\centering
		\includegraphics[scale=0.45]{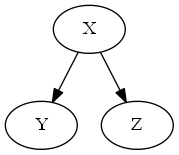}
		\subcaption{$M_1: Y \condindep Z|X$ \label{fig:pgms_a}}
	\end{minipage}
	\begin{minipage}[b]{0.49\linewidth}
		\centering
		\includegraphics[scale=0.45]{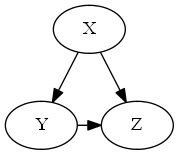}
		\subcaption{$M_2: Y \not\condindep Z|X$ \label{fig:pgms_b}}
	\end{minipage}
	\caption{Simple probabilistic graphical models. In (a) model $M_1$, where $Y$ is conditionally independent of $Z$ 
	given $X$, in (b) model $M_2$, where $Y$ is \emph{not} conditionally independent of $Z$ given $X$.\label{fig:pgms}}
\end{figure}

The two probabilistic graphical models ($M_1$ and $M_2$) are shown in Figure~\ref{fig:pgms}, where all the three variables $X$, $Y$ and $Z$ assume values in $\{1,2,3\}$: in the first model (Figure~\ref{fig:pgms_a}), the hypothesis of independence $H:Y \condindep Z|X$ is \emph{true}, while in the second model (Figure~\ref{fig:pgms_b}), the same hypothesis is \emph{false}. The synthetic conditional probability distribution tables (CPTs) used to generate the samples are given in Appendix~\ref{sec:appendix}.

We calculate the intervals for the \emph{e-values}, and compare them, for the hypothesis $H$, of conditional independence, for both models: $Ev_{M_1}\left(H\right)$ and $Ev_{M_2}\left(H\right)$. The complexity hypothesis $H$ can be decomposed in elementary components:
\begin{align*}
H_1: Y \condindep Z|X=1\\
H_2: Y \condindep Z|X=2\\
H_3: Y \condindep Z|X=3
\end{align*}
\begin{table}[h]
	\caption{\footnotesize Contingency tables of $Y$ and $Z$ for a given the value of $X$ for 5,000 
	random samples. In (a),(c),(e) samples from model $M_1$ (Figure~\ref{fig:pgms_a}) for $X=1$,$2$ 
	and $3$, respectivelly, in (b),(d),(f) samples from model $M_2$ (Figure~\ref{fig:pgms_b}) for 
	$X=1$,$2$ and $3$, respectivelly\label{tab:conttableex}}
	\begin{minipage}[b]{0.49\linewidth}
	\centering\footnotesize
		\subcaption{\footnotesize Model $M_1$ (for $X=1$)\label{tab:conttableexm1x1}}
		\begin{tabular}{l || c | c | c | c |}
			\hline\noalign{\smallskip}
				& $Z=1$ &  $Z=2$ & $Z=3$ & \\
			\noalign{\smallskip}\hline\noalign{\smallskip}
			$Y=1$ & 241 & 187 & 44 & 472\\
			$Y=2$ & 139 & 130 & 30 & 299\\
			$Y=3$ & 364 & 302 & 70 & 736\\
			\hline
			      & 744 & 619 & 144 & 1507\\
			\noalign{\smallskip}\hline
		\end{tabular}
	\end{minipage}
	\begin{minipage}[b]{0.49\linewidth}
	\centering\footnotesize
		\subcaption{\footnotesize Model $M_2$ (for $X=1$)\label{tab:conttableexm2x1}}
		\begin{tabular}{l || c | c | c | c |}
			\hline\noalign{\smallskip}
				& $Z=1$ &  $Z=2$ & $Z=3$ & \\
			\noalign{\smallskip}\hline\noalign{\smallskip}
			$Y=1$ & 228 & 179 & 39  & 446\\
			$Y=2$ & 25  & 33  & 211 & 269\\
			$Y=3$ & 482 & 75  & 208 & 765\\
			\hline
			      & 735 & 287 & 458 & 1048\\
			\noalign{\smallskip}\hline
		\end{tabular}
	\end{minipage}\\
	\begin{minipage}[b]{0.49\linewidth}
	\centering\footnotesize
		\subcaption{\footnotesize Model $M_1$ (for $X=2$)\label{tab:conttableexm1x2}}
		\begin{tabular}{l || c | c | c | c |}
			\hline\noalign{\smallskip}
				& $Z=1$ &  $Z=2$ & $Z=3$ & \\
			\noalign{\smallskip}\hline\noalign{\smallskip}
			$Y=1$ & 42 & 41  & 323 & 406\\
			$Y=2$ & 39 & 41  & 341 & 421\\
			$Y=3$ & 15 & 21  & 171 & 207\\
			\hline
			      & 96 & 103 & 835 & 1034\\
			\noalign{\smallskip}\hline
		\end{tabular}
	\end{minipage}
	\begin{minipage}[b]{0.49\linewidth}
	\centering\footnotesize
		\subcaption{\footnotesize Model $M_2$ (for $X=2$)\label{tab:conttableexm2x2}}
		\begin{tabular}{l || c | c | c | c |}
			\hline\noalign{\smallskip}
				& $Z=1$ &  $Z=2$ & $Z=3$ & \\
			\noalign{\smallskip}\hline\noalign{\smallskip}
			$Y=1$ & 77  & 85  & 248 & 410\\
			$Y=2$ & 165 & 135 & 120 & 420\\
			$Y=3$ & 188 & 21  &  24 & 233\\
			\hline
			      & 430 & 241 & 392 & 1036\\
			\noalign{\smallskip}\hline
		\end{tabular}
	\end{minipage}	
	\begin{minipage}[b]{0.49\linewidth}
	\centering\footnotesize
		\subcaption{\footnotesize Model $M_1$ (for $X=3$)\label{tab:conttableexm1x3}}
		\begin{tabular}{l || c | c | c | c |}
			\hline\noalign{\smallskip}
				& $Z=1$ &  $Z=2$ & $Z=3$ & \\
			\noalign{\smallskip}\hline\noalign{\smallskip}
			$Y=1$ & 282  &  35 & 151 & 468\\
			$Y=2$ & 131  &  37 &  79 & 247\\
			$Y=3$ & 1055 & 143 & 546 & 1744\\
			\hline
			      & 1468 & 215 & 776 & 2459\\
			\noalign{\smallskip}\hline
		\end{tabular}
	\end{minipage}
	\begin{minipage}[b]{0.49\linewidth}
	\centering\footnotesize
		\subcaption{\footnotesize Model $M_2$ (for $X=3$)\label{tab:conttableexm2x3}}
		\begin{tabular}{l || c | c | c | c |}
			\hline\noalign{\smallskip}
				& $Z=1$ &  $Z=2$ & $Z=3$ & \\
			\noalign{\smallskip}\hline\noalign{\smallskip}
			$Y=1$ & 40  & 87   & 354 & 481\\
			$Y=2$ & 119 & 104  &  27 & 250\\
			$Y=3$ & 305 & 1049 & 372 & 1726\\
			\hline
			      & 464 & 1240 & 753 & 2457\\
			\noalign{\smallskip}\hline
		\end{tabular}
	\end{minipage}	
\end{table}

For each model, $5,000$ random observation have been generated, the contingency table of $Y$ and $Z$ for each value of $X$ are shown in Table~\ref{tab:conttableex}. The hyperparameters of the prior distribution were all set to 1 , the priori is then equivalent to a uniform distribution (from Equation~\ref{eq:priordist}):
\begin{align*}
\alpha_1=\alpha_2=\alpha_3=[1,1,1] \\
f\left(\theta_1|\alpha_1\right)= f\left(\theta_3|\alpha_3\right)=f\left(\theta_3|\alpha_3\right)=1
\end{align*}
The posterior distribution, found using Equations~\ref{eq:multinomialdist} and~\ref{eq:posteriordist}, is then:
\begin{align*}
f_n\left(\theta_1\right) \propto \prod_{y=1,z=1}^{3,3}\theta_{yz1}^{n_{yz1}},
f_n\left(\theta_2\right) \propto \prod_{y=1,z=1}^{3,3}\theta_{yz2}^{n_{yz2}},
f_n\left(\theta_3\right) \propto \prod_{y=1,z=1}^{3,3}\theta_{yz3}^{n_{yz3}}
\end{align*}
For example, for the given contingency table for Model $M_1$, when $X=2$ (Table~\ref{tab:conttableexm1x2}) the posterior distribution is:
\begin{align*}
f_n\left(\theta_2\right) \propto 
\theta_{112}^{42}\cdot
\theta_{122}^{41}\cdot
\theta_{132}^{323}\cdot
\theta_{212}^{39}\cdot
\theta_{222}^{41}\cdot
\theta_{232}^{341}\cdot
\theta_{312}^{15}\cdot
\theta_{322}^{21}\cdot
\theta_{332}^{171}
\end{align*}
And the point of highest density, for this example, under the hypothesis of independence (Equations~\ref{eq:pointmaxdensityH} and~\ref{eq:maxdensityH}) was found to be:
\begin{align*}
\theta_2^* \approx \left[0.036,0.039,0.317,0.038,0.041,0.329,0.019,0.020,0.162\right]
\end{align*}
The truth function and the evidence supporting the hypothesis of independence given $X=2$ (hypothesis $H_2$) for model $M_1$, as given in Equations~\ref{eq:regiontruthfunc} and~\ref{eq:evalue}, are:
\begin{align*}
R_2(f) = \{\theta_2 \in \Theta_2|f_n\left(\theta_2\right)\leq f\} \nonumber \\
W_2(f) = \int_{R_2(f)}f_n\left(\theta_2\right)d\theta_2 \nonumber \\
Ev_{M_1}(H_2)=W_2(f_{n}(\theta_2^*))
\end{align*}
We used methods of numerical integration to find the e-value of the elementary components of hypothesis $H$ ($H_1$,$H_2$ and $H_3$), the results for each model are given bellow.

\begin{figure}[H]
	\begin{minipage}[b]{0.49\linewidth}
	\centering
		\includegraphics[scale=0.30]{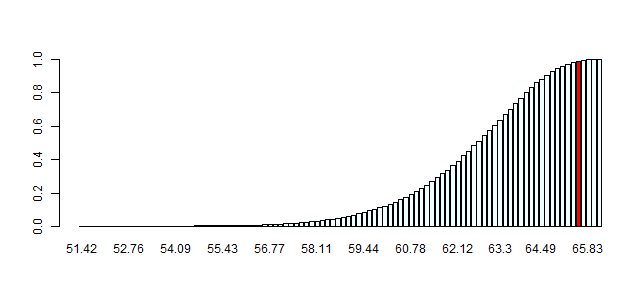}
		\subcaption{$W_1$ for Model $M_1$, $f_n(\theta_1^*)$ in red.\newline
		Horizontal Discretization.\label{fig:cumhistM1X1}}
	\end{minipage}
	\begin{minipage}[b]{0.49\linewidth}
	\centering
		\includegraphics[scale=0.30]{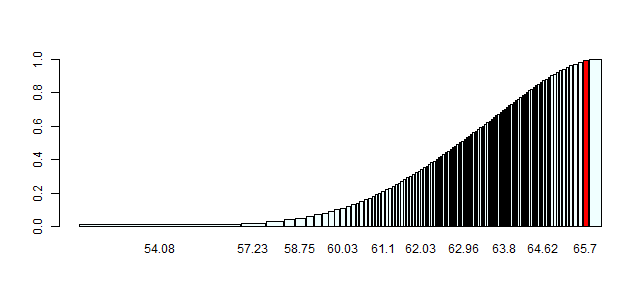}
		\subcaption{$W_1$ for Model $M_1$, $f_n(\theta_1^*)$  in red.\newline
		Vertical Discretization.\label{fig:cumhistM1X1V}}
	\end{minipage}\\
	\begin{minipage}[b]{0.49\linewidth}
	\centering
		\includegraphics[scale=0.30]{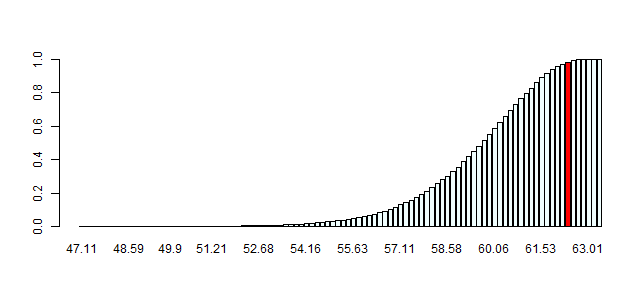}
		\subcaption{$W_2$, for Model $M_1$, $f_n(\theta_2^*)$  in red.\newline
		Horizontal Discretization.\label{fig:cumhistM1X2}}
	\end{minipage}
	\begin{minipage}[b]{0.49\linewidth}
	\centering
		\includegraphics[scale=0.30]{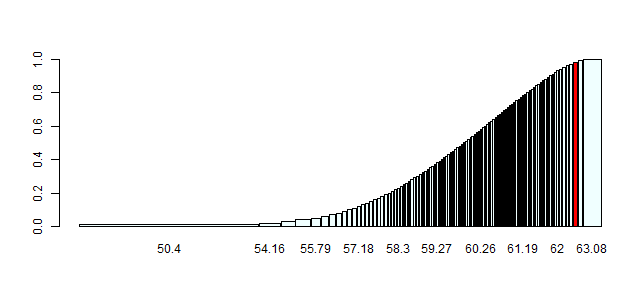}
		\subcaption{$W_2$, for Model $M_1$, $f_n(\theta_2^*)$  in red.\newline
		Vertical Discretization.\label{fig:cumhistM1X2V}}
	\end{minipage}
	\begin{minipage}[b]{0.49\linewidth}
	\centering
		\includegraphics[scale=0.30]{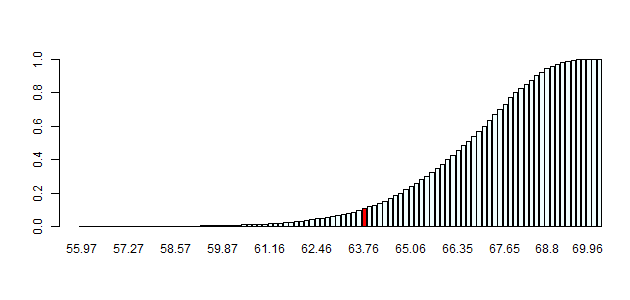}
		\subcaption{$W_3$, for Model $M_1$, $f_n(\theta_3^*)$  in red.\newline
		Horizontal Discretization.\label{fig:cumhistM1X3}}
	\end{minipage}
	\begin{minipage}[b]{0.49\linewidth}
	\centering
		\includegraphics[scale=0.30]{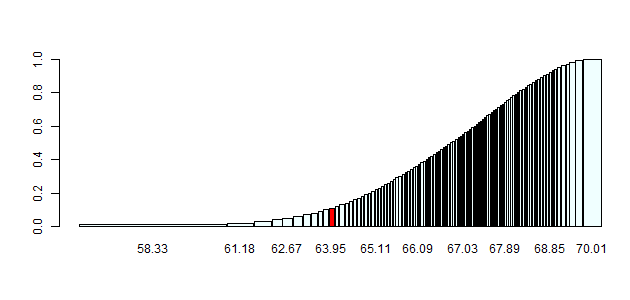}
		\subcaption{$W_3$, for Model $M_1$, $f_n(\theta_3^*)$  in red.\newline
		Vertical Discretization.\label{fig:cumhistM1X3V}}
	\end{minipage}
	\caption{Histogram with 100 bins of the truth functions for the Model $M_1$ (Figure~\ref{fig:pgms_a}), 
	for each value of $X$. In red, the maximum posterior density under the respective elementary component 
	($H_1$, $H_2$ and $H_3$) of the hypothesis of conditional independence $H$, for both horizontal and 
	vertical discretization procedures.
	\label{fig:W1W2W3forM1}}
\end{figure}

\textit{E-values} found using \emph{horizontal} discretization:
\begin{align*}
Ev_{M_1}(H_1)=0.9878, Ev_{M_1}(H_2)=0.9806 \text{ and } Ev_{M_1}(H_3)=0.1066 \\
Ev_{M_2}(H_1)=0.0004, Ev_{M_2}(H_2)=0.0006 \text{ and } Ev_{M_2}(H_3)=0.0004 
\end{align*}
, and the \textit{e-values} found using \emph{vertical} discretization:
\begin{align*}
Ev_{M_1}(H_1)=0.99, Ev_{M_1}(H_2)=0.98 \text{ and } Ev_{M_1}(H_3)=0.11\\
Ev_{M_2}(H_1)=0.01, Ev_{M_2}(H_2)=0.01 \text{ and } Ev_{M_2}(H_3)=0.01 
\end{align*}
Figure~\ref{fig:W1W2W3forM1} shows the histogram of the Truth functions $W_1$, $W_2$ and $W_3$ for the Model $M_1$
($Y$ and $Z$ are conditionally independent given $X$). In Figures~\ref{fig:cumhistM1X1},~\ref{fig:cumhistM1X2}~and \ref{fig:cumhistM1X3}, $100$ bins are uniformly distributed over the $x$ axis (using the empirical values of $\min f_n(\theta_x)$ and $\max f_n(\theta_x)$). In Figures~\ref{fig:cumhistM1X1V},~\ref{fig:cumhistM1X2V}~and \ref{fig:cumhistM1X3V}, $100$ bins are uniformly distributed over the $y$ axis (each bin represents an increase in $1\%$ in density from the previous bin). Notice that the functions $W_x$ evaluated at the maximum posterior density over the respective hypothesis $f_n(\theta_x^*)$, in red, correspond to the e-values found (e.g., $W_3(f(\theta_3^*)) \approx 0.1066$, for the horizontal discretization in Figure~\ref{fig:cumhistM1X3}).

\begin{figure}[H]
	\begin{minipage}[b]{0.49\linewidth}
		\centering
		\includegraphics[scale=0.30]{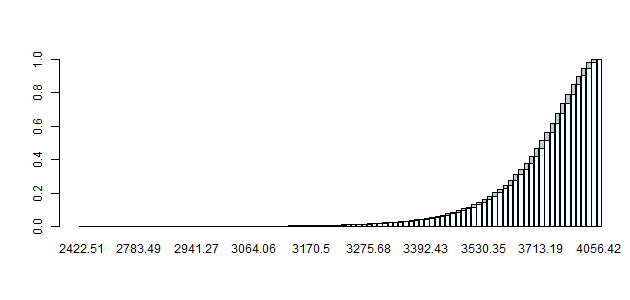}
		\subcaption{$W_1\otimes W_2$ for Model $M_1$.\\
		Horizontal Discretization.\label{fig:convW1W2M1}}
	\end{minipage}
	\begin{minipage}[b]{0.49\linewidth}
		\centering
		\includegraphics[scale=0.30]{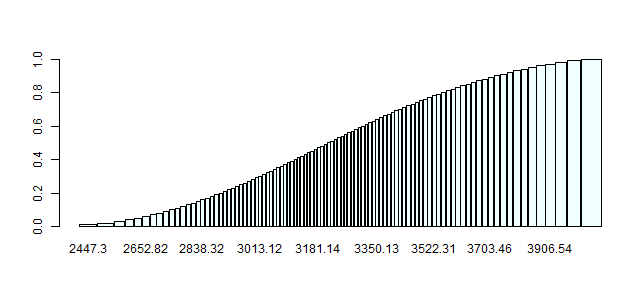}
		\subcaption{$W_1\otimes W_2$ for Model $M_1$.\\
		Vertical Discretization.\label{fig:convW1W2M1V}}
	\end{minipage}\\
	\begin{minipage}[b]{0.49\linewidth}
		\centering
		\includegraphics[scale=0.30]{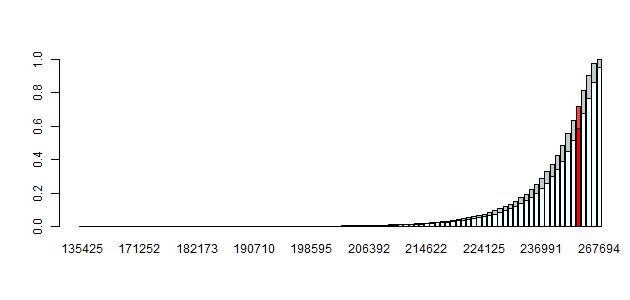}
		\subcaption{$W_{1}\otimes W_{2}\otimes W_3$ for Model $M_1$.\\
		Horizontal Discretization.\label{fig:convW1W2W3M1}}
	\end{minipage}
	\begin{minipage}[b]{0.49\linewidth}
		\centering
		\includegraphics[scale=0.30]{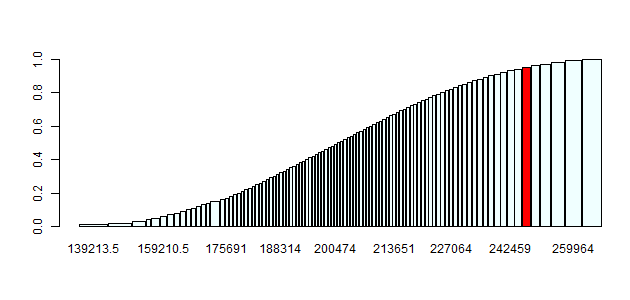}
		\subcaption{$W_{1}\otimes W_{2}\otimes W_3$ for Model $M_1$.\\
		Vertical Discretization.\label{fig:convW1W2W3M1V}}
	\end{minipage}\\
	\caption{Histogram with 100 bins of the resulting convolutions for Model $M_1$: 
	(a) $W_1 \otimes W_2$ with horizontal discretization;
	(b) $W_1 \otimes W_2$ with vertical discretization;
	(c) $W_1 \otimes W_2 \otimes W_3$ with horizontal discretization;
	(d) $W_1 \otimes W_2 \otimes W_3$ with vertical discretization.
	In red in (c) and (d), the bin representing the product of maximum posterior density under
	the elementary components ($H_1$, $H_2$ and $H_3$) of the hypothesis of conditional independence $H$
	for model $M_1$.
	\label{fig:convM1}}
\end{figure}

\begin{figure}[H]
	\begin{minipage}[b]{0.49\linewidth}
		\centering
		\includegraphics[scale=0.30]{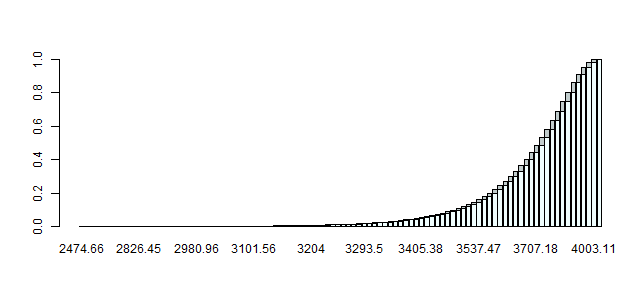}
		\subcaption{$W_1\otimes W_2$ for Model $M_2$.\\
		Horizontal Discretization.\label{fig:convW1W2M2}}
	\end{minipage}
	\begin{minipage}[b]{0.49\linewidth}
		\centering
		\includegraphics[scale=0.30]{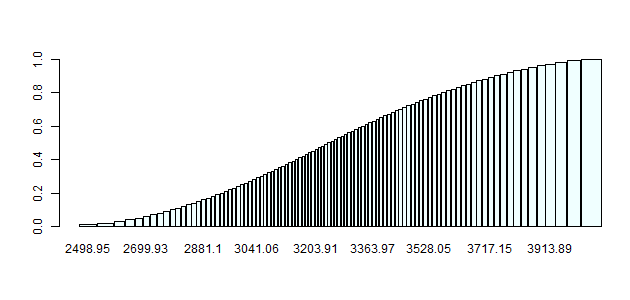}
		\subcaption{$W_1\otimes W_2$ for Model $M_2$.\\
		Vertical Discretization.\label{fig:convW1W2M2V}}
	\end{minipage}\\
	\begin{minipage}[b]{0.49\linewidth}
		\centering
		\includegraphics[scale=0.30]{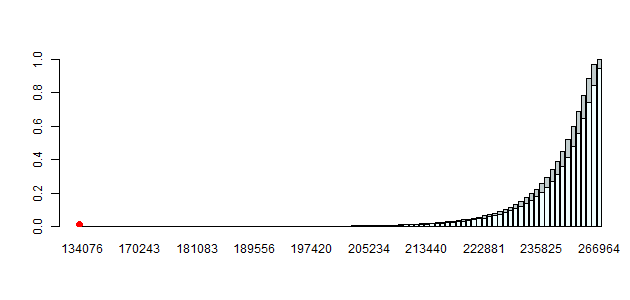}
		\subcaption{$W_{1}\otimes W_{2}\otimes W_3$ for Model $M_2$.\\
		Horizontal Discretization.\label{fig:convW1W2W3M2}}
	\end{minipage}
	\begin{minipage}[b]{0.49\linewidth}
		\centering
		\includegraphics[scale=0.30]{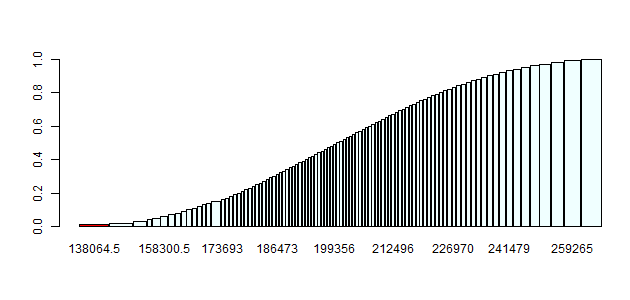}
		\subcaption{$W_{1}\otimes W_{2}\otimes W_3$ for Model $M_2$.\\
		Vertical Discretization.\label{fig:convW1W2W3M2V}}
	\end{minipage}
	\caption{Histogram with 100 bins of the resulting convolutions for Model $M_2$: 
	(a) $W_1 \otimes W_2$ with horizontal discretization;
	(b) $W_1 \otimes W_2$ with vertical discretization;
	(c) $W_1 \otimes W_2 \otimes W_3$ with horizontal discretization;
	(d) $W_1 \otimes W_2 \otimes W_3$ with vertical discretization.
	In red in (c) and (d), the bin representing the product of maximum posterior density under
	the elementary components ($H_1$, $H_2$ and $H_3$) of the hypothesis of conditional independence $H$
	for model $M_2$.
	\label{fig:convM2}}
\end{figure}

The evidence supporting the hypothesis of conditional independence $H$, as in Equation~\ref{eq:Hevalue}, for each model, will be:
\begin{align*}
Ev(H)=W_1\otimes W_2 \otimes W_3 \left(f_{n}(\theta_1^*)\cdot f_{n}(\theta_2^*)\cdot f_{n}(\theta_3^*)\right)
\end{align*}
The convolution has commutative property, therefore the order of the convolutions is irrelevant: 
\begin{align*}
W_1\otimes W_2 \otimes W_3 (f)=W_3\otimes W_2 \otimes W_1 (f)
\end{align*}
, using the algorithm for numerical convolution described in Algorithm~\ref{alg:convolution} we found the convolution of the truth functions $W_1$ and $W_2$, resulting in a cumulative function ($W_{12}$) with $10,000$ bins ($100^2$ bins). We, then, performed the condensation procedures described in Algorithms~\ref{alg:condensation}~\ref{alg:verticalcondensation}, reducing the cumulative distribution to 100 bins, with lower and upper bounds ($W^l_{12}$ and $W^u_{12}$) for the horizontal condensation. The results are shown in Figures~\ref{fig:convW1W2M1} and~\ref{fig:convW1W2M1V} for Model $M_1$ (horizontal and vertical condensations, respectively), and, ~\ref{fig:convW1W2M2} and~\ref{fig:convW1W2M2V} for model $M_2$.

The convolution of $W_{12}$ and $W_{3}$ was, then, performed, followed by condensation. The results, are shown in Figures~\ref{fig:convW1W2W3M1} and~\ref{fig:convW1W2W3M1V} (model $M_1$), and~\ref{fig:convW1W2W3M2} and~\ref{fig:convW1W2W3M2V} (model $M_2$). 

The \textit{e-values} supporting the hypothesis of conditional independence for both models are given bellow.

The intervals for the \textit{e-values} found using horizontal discretization and condensation were:
\begin{align*}
Ev_{M_1}(H)=[0.587427,0.718561]\\
Ev_{M_2}(H)=[8\cdot 10^{-12}, 6.416\cdot 10^{-9}]
\end{align*}
, and the \textit{e-values} found using vertical discretization and condensation were:
\begin{align*}
Ev_{M_1}(H)= 0.95\\
Ev_{M_2}(H)=0.01
\end{align*}
These results show strong evidence supporting the hypothesis of conditional independence between $Y$ and $Z$ given $X$ for the model $M_1$ (using both discretization/condensation procedures). And no evidence supporting the same hypothesis for the second model. This result is very relevant and promising as a motivation for further studies of the use of FBST as a CI test for the structure learning of graphical models.

\section{Conclusion and Future Work}
\label{sec:conclusion}
This paper gives the framework to perform tests of conditional independence for discrete datasets using the Full Bayesian Significance Test (FBST). A simple example of application of this test to learn the structure of a directed acyclic graph is given using two different models. The result found in this paper suggests that FBST should be considered as a good alternative to perform CI tests for the task of learning structures of probabilistic graphical models from data.

Future researches include the use of FBST in an algorithm to learn structures of graphs with larger number of variables; the increase in performance of the mathematical methods used to calculate the e-values (as learning DAG structures from data requires an exponential number of CI tests to be performed, each CI test needs to be performed faster); and an empirical evaluation of the threshold for e-values in order to define conditional independence versus dependence, by minimizing a linear combination of errors of type I and II (incorrect rejection of true hypothesis of conditional independence and failure to reject a false hypothesis of conditional independence).

\clearpage
\appendix
\section{Appendix}
\label{sec:appendix}
\begin{table}[H]
	\caption{\footnotesize Conditional probability distribution tables. In (a) the distribution of $X$, 
	in (b) conditional distribution of $Y$, given $X$, 
	in (c) conditional distribution of $Z$, given $X$.}
	\begin{minipage}[b]{0.27\linewidth}
	\centering\footnotesize
		\subcaption{\footnotesize CPT of $X$}
		\begin{tabular}{l || c }
			\hline\noalign{\smallskip}
				$X$ &  p($X$)\\
			\noalign{\smallskip}\hline\noalign{\smallskip}
				1 & 0.3\\
				2 & 0.2\\
				3 & 0.5\\
			\noalign{\smallskip}\hline
		\end{tabular}
	\end{minipage}
	\begin{minipage}[b]{0.67\linewidth}
	\centering\footnotesize
		\subcaption{\footnotesize CPT of $Y$ given $X$}
		\begin{tabular}{l || c |  c | c | c }
			\hline\noalign{\smallskip}
				$Y$ &  p($Y|X$=1) &  p($Y|X$=2) &  p($Y|X$=3)\\
			\noalign{\smallskip}\hline\noalign{\smallskip}
				1 & 0.3 & 0.4 & 0.2\\
				2 & 0.2 & 0.4 & 0.1\\
				3 & 0.5 & 0.2 & 0.7\\
			\noalign{\smallskip}\hline
		\end{tabular}
	\end{minipage}\\
	\begin{minipage}[b]{1.0\linewidth}
	\centering\footnotesize
		\subcaption{\footnotesize CPT of $Z$ given $X$}
		\begin{tabular}{l || c |  c | c | c }
			\hline\noalign{\smallskip}
				$Z$ &  p($Z|X$=1) &  p($Z|X$=2) &  p($Z|X$=3)\\
			\noalign{\smallskip}\hline\noalign{\smallskip}
				1 & 0.5 & 0.1 & 0.6\\
				2 & 0.4 & 0.1 & 0.1\\
				3 & 0.1 & 0.8 & 0.3\\
			\noalign{\smallskip}\hline
		\end{tabular}
	\end{minipage}
\end{table}

\begin{table}[H]
	\centering\footnotesize
	\caption{\footnotesize Conditional probability distribution table of $Z$, given $X$ \& $Y$.}
	\begin{tabular}{l || c | c | c | c | c }
		\hline\noalign{\smallskip}
			$Z$ &  p($Z|X$=1,$Y$=1) &  p($Z|X$=1,$Y$=2) &  p($Z|X$=1,$Y$=3)\\
			\noalign{\smallskip}\hline\noalign{\smallskip}
				1 & 0.5 & 0.1 & 0.6\\
				2 & 0.4 & 0.1 & 0.1\\
				3 & 0.1 & 0.8 & 0.3\\
			\noalign{\smallskip}\hline
	\end{tabular}\\
	\begin{tabular}{l || c | c | c | c | c }
		\hline\noalign{\smallskip}
			$Z$ &  p($Z|X$=2,$Y$=1) &  p($Z|X$=2,$Y$=2) &  p($Z|X$=2,$Y$=3)\\
			\noalign{\smallskip}\hline\noalign{\smallskip}
				1 & 0.2 & 0.4 & 0.8\\
				2 & 0.2 & 0.3 & 0.1\\
				3 & 0.6 & 0.3 & 0.1\\
			\noalign{\smallskip}\hline
	\end{tabular}\\
	\begin{tabular}{l || c | c | c | c }
		\hline\noalign{\smallskip}
			$Z$ &  p($Z|X$=3,$Y$=1) &  p($Z|X$=3,$Y$=2) &  p($Z|X$=3,$Y$=3)\\
		\noalign{\smallskip}\hline\noalign{\smallskip}
			1 & 0.1 & 0.5 & 0.2\\
			2 & 0.2 & 0.4 & 0.6\\
			3 & 0.7 & 0.1 & 0.2\\
		\noalign{\smallskip}\hline
	\end{tabular}
\end{table}

\end{document}